\documentclass[doublecol]{epl2}
\usepackage{amssymb}
\usepackage{graphicx}
\usepackage{booktabs}
\usepackage{tabularx}
\usepackage{array}
\usepackage{lscape}
\usepackage{longtable}
\usepackage{multirow}
\usepackage{amsmath}
\usepackage{colortbl}
\usepackage{algorithm}
\usepackage{algorithmic}

\newtheorem{definition}{Definition}
\title{The time series forecasting: from the aspect of network}
\shorttitle{title} 

\author{S. Chen\inst{1} \and X. Lan\inst{1} \and Y. Hu\inst{2} \and Q. Liu\inst{3,4} \and Y. Deng\inst{1}}
\shortauthor{S. Chen \etal}

\institute{
  \inst{1} School of Computer and Information Science, Southwest University - Chongqing 400715, China\\
  \inst{2} Institute of Business Intelligence and Knowledge Discovery, Guangdong University of Foreign Studies, Guangzhou 510006, China\\
  \inst{3} Department of Biomedical Informatics, Medical Center, Vanderbilt University, Nashiville, 37235, USA\\
  \inst{4} School of Life Sciences and Biotechnology, Shanghai Jiao Tong University, Shanghai, 200030, China
}



\abstract{
Forecasting can estimate the statement of events according to the historical data and it is considerably important in many disciplines. At present, time series models have been utilized to solve forecasting problems in various domains. In general, researchers use curve fitting and parameter estimation methods (moment estimation, maximum likelihood estimation and least square method) to forecast. In this paper, a new sight is given to the forecasting and a completely different method is proposed to forecast time series. Inspired by the visibility graph and link prediction, this letter converts time series into network and then finds the nodes which are mostly likelihood to link with the predicted node. Finally, the predicted value will be obtained according to the state of the link. The TAIEX data set is used in the case study to illustrate that the proposed method is effectiveness. Compared with ARIMA model, the proposed shows a good forecasting performance when there is a small amount of data.}

\begin{document}

\maketitle

\section{Introduction}
Forecasting estimates the statement of events in the future according to the historical data and it is considerably important in many disciplines, such as finance, meteorology, industry and so forth. At present, abundant time series models have been utilized to solve forecasting problems.

For instance, exponential smoothing methods were introduced in the 1950s by the works of Brown and whereafter these methods got a great development. In general, the smoothing parameters are restricted to the range 0 to 1 but the usual intervals may produce non-invertible models. Autoregressive and Moving Average (ARMA) model is an important method to study time series. The concept of autoregressive (AR) and moving average (MA) models was formulated by the works of Yule, Slutsky, Walker and Yaglom. Autoregressive integrated moving average (ARIMA) model \cite{box2013time} is based on the ARMA model. The difference is that ARIMA model converts non-stationary time series into stationary time series before adopting ARMA model. ARMA and ARIMA model are widely used to predict linear time series. In order to predict non-linear time series, some other models are proposed. Artificial neural network (ANN) is useful for nonlinear processes, so it is applied in the area of forecasting \cite{hippert2001neural,zhang1998forecasting}. In artificial neural network, the inputs or variables get filter through one or more hidden layers and the intermediate output is related to the final output. Besides exponential smoothing methods, ARMA model, ARIMA model and ANN model, there are some other methods to study time series, such as autoregressive conditional heteroscedastic (ARCH) model, generalized autoregressive conditional heteroscedastic (GARCH) model, long memory models, structural models and so forth. These methods have their respective characteristics.

In this letter, a new sight is given to the forecasting. Different from the existed methods, the proposed method in this letter forecasts the time series according to the network structure. Inspired by the visibility graph and the link prediction, this letter converts time series into network based on the visibility graph and finds the relationship between the predicted node and other nodes based on the link prediction. Link prediction can estimate the likelihood of the existence of a link between two nodes and commonly, if two nodes are more similar, they are more likely to be connected. In other words, we can find which nodes the predicted node will link with by using link prediction. And then, a method which can covert network into time series is proposed to calculate the value of the predicted node according to the state of the link.

This letter presents a new aspect of forecasting. The proposed method is compared with classic ARIMA model in forecasting stock price. We use time series of different length as training set to forecast stock price. The experimental results show that when the number of data in training set is small, the forecasting of the proposed method is better than ARIMA model. In general, the forecasting effect of the proposed method is favorable.

\section{Preliminary knowledge}

\subsection{The visibility graph}
Complex networks are widely used in many disciplines \cite{watts1998collective,barabasi1999emergence,newman2009,liu2009,vidal2011,wei2013identifiy,donges2013testing,wei2013fractal,chenduanbing2013ClusterRank,liujianguo2013ranking}. Recently, a new tool called the visibility graph is proposed to transform time series into network which builds a bridge between time series and network. The visibility graph method was proposed by L. Lacasa et al. in 2008 \cite{lacasa2008time}. In the visibility graph, the values of time series are plotted by using vertical bars. A vertical bar links with others and the visibility criteria is established in the literature \cite{lacasa2008time} as follows:

\begin{definition}
Two data value $({t_1},{y_1})$ and $({t_2},{y_2})$ have visibility, if any other value $({t_3},{y_3})$ is placed between them fulfills:
\begin{equation}
{y_3} < {y_2} + ({y_1} - {y_2})\frac{{{t_2} - {t_3}}}{{{t_2} - {t_1}}}
\end{equation}
\end{definition}

\begin{figure}[!t]
\centering
\includegraphics[scale=0.4]{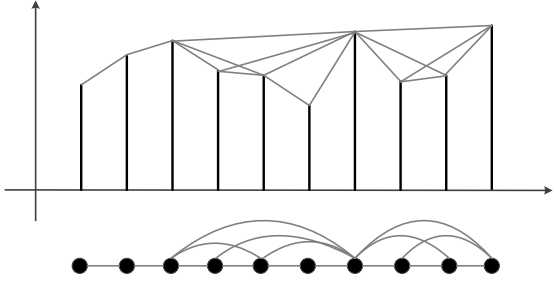}
\caption{\emph{The histogram shows a time series with 10 data values, and according to the visibility algorithm, the associated graph is obtained. In the histogram, if a bar can be seen from the top of considered one, they will be linked. If two bar are linked in the histogram, the nodes which represent them will be linked in the associated graph.}}
\label{visibilityexample}
\end{figure}

The associated graph derived from a time series has the following properties:
1)Connected: a node can see its nearest neighbors.
2)Undirected: the associated graph extracted from a time series is undirected.
3)Invariant under affine transformations of the series data: Although rescale both horizontal and vertical axes, the visibility criterion is invariant.

Until now, the visibility graph has been applied in many discipline to analysis the time series \cite{telesca2012analysis,lacasa2009visibility,mehraban2013coupling,ahmadlou2012improved,telesca2013investigating,donner2012visibility,pierini2012visibility}.

\subsection{Link prediction}
Recently, link prediction is paid much attention in complex networks. It can find the links which likely appear in the network. The literature \cite{liu2010link} proposes an efficient method to deal with the problem of link prediction. This method is based on local random walk and exploits the possible links by calculating the similarity between two nodes. If two nodes are more similar, they are more likely to be connected. The main process is described as follows.

Supposing there is an undirected network $G(V,E)$, where $V$ is the set of nodes and $E$ is the set of links. For each pair of nodes, $x,y \in V$, a score ${S_{xy}}$ denotes the similarity between node $x$ and node $y$.

Random walk process can be described by the transition probability matrix $R$. ${R_{xy}} = {a_{xy}}/{k_x}$ presents the probability that a walker who stays at node $x$ will move to node $y$ in the next step, where if the node $x$ links with the node $y$,${a_{xy}} = 1$; otherwise, ${a_{xy}} = 0$, and ${k_x}$ is the degree of the node $x$. The following equation can measure the probability that a walker starting from node $x$ locates at other nodes after $t$ steps.

\begin{equation}
\label{probability}
{\overrightarrow \pi  _x}(t) = {R^T}{\overrightarrow \pi  _x}(t - 1)
\end{equation}

where ${\overrightarrow \pi  _x}(0)$ is an $N \times 1$ vector with the $x$-th element equals to 1 and others to 0. Then, the similarity between node $x$ and node $y$ is defined as follows:

\begin{equation}
\label{lrw}
S_{xy}^{LRW}(t) = \frac{{{k_x}}}{{2M}}{\pi _{xy}}(t) + \frac{{{k_y}}}{{2M}}{\pi _{yx}}(t)
\end{equation}
where $M$ is the number of links and there exists ${S_{xy}} = {S_{yx}}$. To let a random walker circulate locally rather than go too far away, a equation is defined as follows which superposes the contribution of each walker.

\begin{equation}
\label{srw}
S_{xy}^{SRW}(t) = \sum\limits_{i = 1}^t {S_{xy}^{LRW}(i)}
\end{equation}

In the literature \cite{liu2010link}, the authors have proved that this link prediction method based on local random walk is efficient and compared with six well-known methods on five real networks, it gives better prediction than the three local similarity indices.

\section{Proposed method}
Inspired by the visibility graph and link prediction, we propose a new method to forecast time series from the aspect of network. The time series is converted into the network to exploit the relationship between the predicted node and other nodes. Then, this network will be converted into time series by our method to calculate the value of the predicted one. The detailed process is described as follows.

Supposing there is a time series $({t_1},{x_1})$, $({t_2},{x_2})$, $ \cdots $, $({t_i},{x_i})$ $ \cdots $, $({t_n},{x_n})$. $x_i$ is the data value and $t_i$ is the time when $x_i$ is observed.

\begin{figure}[!t]
\centering
\includegraphics[scale=0.5]{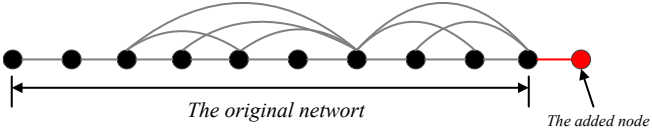}
\caption{\emph{Supposing the original network is like fig.\ref{visibilityexample} showing. The added node denotes the datum which needs to be predicted. It must link with the last datum in the time series.}}
\label{add}
\end{figure}

\emph{Step1}: The time series is converted into network by using the visibility graph algorithm. The node $y$ which denotes the data needing to be predicted. The node $y$ needs be added to the network. Every node see at least its nearest neighbors, so the predicted node $y$ should link with the node which denotes the last data in the time series. Then, a new network will be obtained. The fig.\ref{add} shows this process.

\emph{Step2}: In the new network, The link prediction method is used. According to the above Eq.\ref{probability}, Eq.\ref{lrw} and Eq.\ref{srw}, the similarity between the added node $y$ and other nodes $x_i$ will be obtained. A latent assumption is that the the link denotes the similarity between two endpoints. If the similarity between two nodes is higher, these two nodes are more likely to be linked. There, a threshold value is set to judge whether two nodes are linked with each other.

\begin{equation}
T = \frac{{\sum\limits_{i = 1}^{n} {{S_{y{x_i}}}} }}{n}
\end{equation}

where $n$ is the number of the nodes in the original network, and ${S_{y{x_i}}}$ is the similarity between the added node $y$ and other nodes $x_i$. To judge whether the nodes $x_i$ link with the added node $y$, we have:

\begin{equation}
{E_{y{x_i}}} = \left\{ \begin{array}{l}
 1{\kern 1pt} {\kern 1pt} {\kern 1pt} {\kern 1pt} {\kern 1pt} {\kern 1pt} {\kern 1pt} {\kern 1pt} {\kern 1pt} {\kern 1pt} {\kern 1pt} {S_{y{x_i}}} > T \\
 0{\kern 1pt} {\kern 1pt} {\kern 1pt} {\kern 1pt} {\kern 1pt} {\kern 1pt} {\kern 1pt} {\kern 1pt} {\kern 1pt} {S_{y{x_i}}} \le T{\kern 1pt}  \\
 \end{array} \right.
\end{equation}

${E_{y{x_i}}}$ is the edge between the node $x_i$ and the added node $y$. It is obvious that ${E_{y{x_i}}}={E_{x{y_i}}}$. This equation means that if ${S_{y{x_i}}} > T$, the added node $y$ will link with the node $x_i$. If ${S_{y{x_i}}} < T$, the added node $y$ will not link with the node $x_i$. Especially, the last node $x_n$ must link with $y$, so ${E_{y{x_n}}}=1$.

After this step, we will know which nodes will link with the added node $y$. The structure of the new network is clear. The value of the node $y$ will be obtained in the next step.

\emph{Step3}: In this step, the network will be converted into the time series again and the value of the added data which needs to be predicted will be obtained.

In the new time series, the order of the data $x_i$ is decided which is same as the order before they are converted into the network. The value of $x_i$ is same as the previous too. The added data $y$ is the last data in the new time series and its value needs to be calculated. This new time series is plotted by using vertical bars. In the histogram, suppose that $x_j$ (${x_j} \in {x_i}$) links with $y$ and $x_j^*$ is the last one that $x_j$ can see on its right side. We have:

\begin{equation}
\label{valueyi}
{y_j} = \frac{{x_j^* - {x_j}}}{{t_j^* - {t_j}}}(t - {t_j}) + {x_j}
\end{equation}

where $t$ is the time when $y$ appears. The value of data $y$ is defined as follows:

\begin{equation}
\label{valuey}
y = \min ({y_j})
\end{equation}

\begin{figure}[!t]
\centering
\includegraphics[scale=0.4]{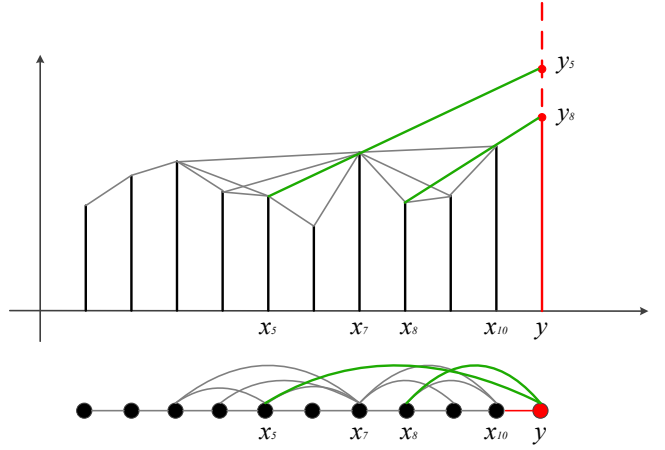}
\caption{\emph{Supposing that according to the step2, we know the node $y$ links with the node $x_5$ and the node $x_8$, respectively. Then convert the network into time series again and plot the new time series by using vertical bars. In the histogram, $x_7$ is the last datum that $x_5$ can see on its right side and $x_{10}$ is the last datum that $x_8$ can see on its right side. According to the Eq.\ref{valuey}, the values of $y_5$ and $y_8$ can be obtained. Because the value of $y_8$ is less than the value of $y_5$, the value $y=y_8$. The predicted value is $y_8$.}}
\label{yvalue}
\end{figure}

Especially, the node $x_n$ is the last one in the original time series. There is no node on $x_n$'s right side, so $x_n$ is not considered in the above calculation. The fig.\ref{yvalue} shows the process of obtaining the value of $y$.

\section{Case Study}
Taiwan stock Exchange Capitalization Weighted Stock Index values in 2012 are used to verify the forecasting performance of the proposed method. In this letter, these data are regarded as a continuous data. The autoregressive integrated moving average (ARIMA) model is used to be compared with the proposed method. To estimate the forecasting accuracy of the proposed method, the root mean square errors (RMSE) is used as a performance measure which is shown as follows:

\begin{equation}
RMSE{\rm{ = }}\sqrt {\frac{{\sum\limits_{t = 1}^n {{\rm{|actual(t) - forecast(t)}}{{\rm{|}}^2}} }}{n}}
\end{equation}

\begin{figure}[!t]
\centering
\includegraphics[scale=0.4]{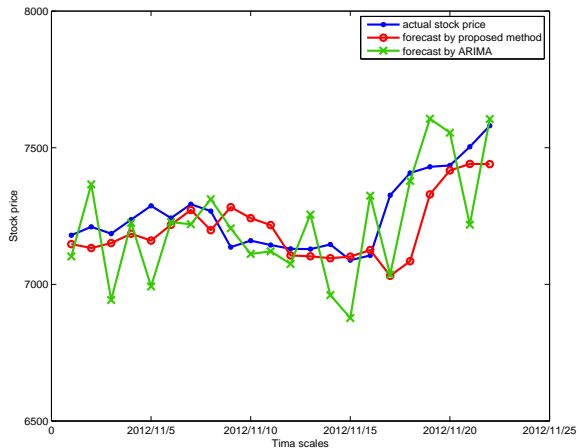}
\caption{\emph{The training data consists of the data of October (21 records). The figure shows the predicted stock price of November obtained by ARIMA and the proposed method, respectively.}}
\label{our1ARIMA1}
\end{figure}

\begin{figure}[!t]
\centering
\includegraphics[scale=0.4]{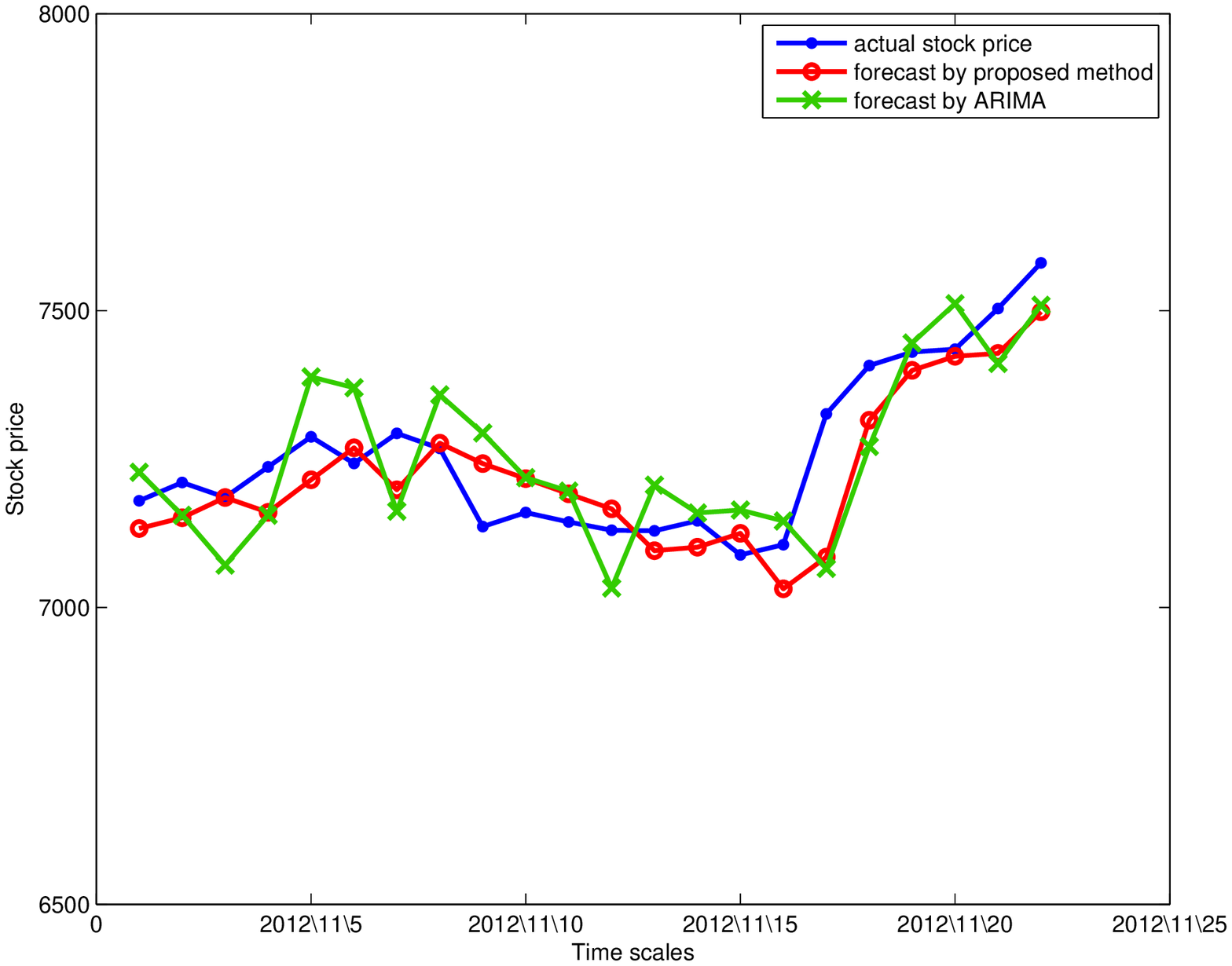}
\caption{\emph{The training data consists of the data of August, September and October (64 records). The figure shows the predicted stock price of November obtained by ARIMA and the proposed method, respectively.}}
\label{our3ARIMA3}
\end{figure}

\begin{figure}[!t]
\centering
\includegraphics[scale=0.4]{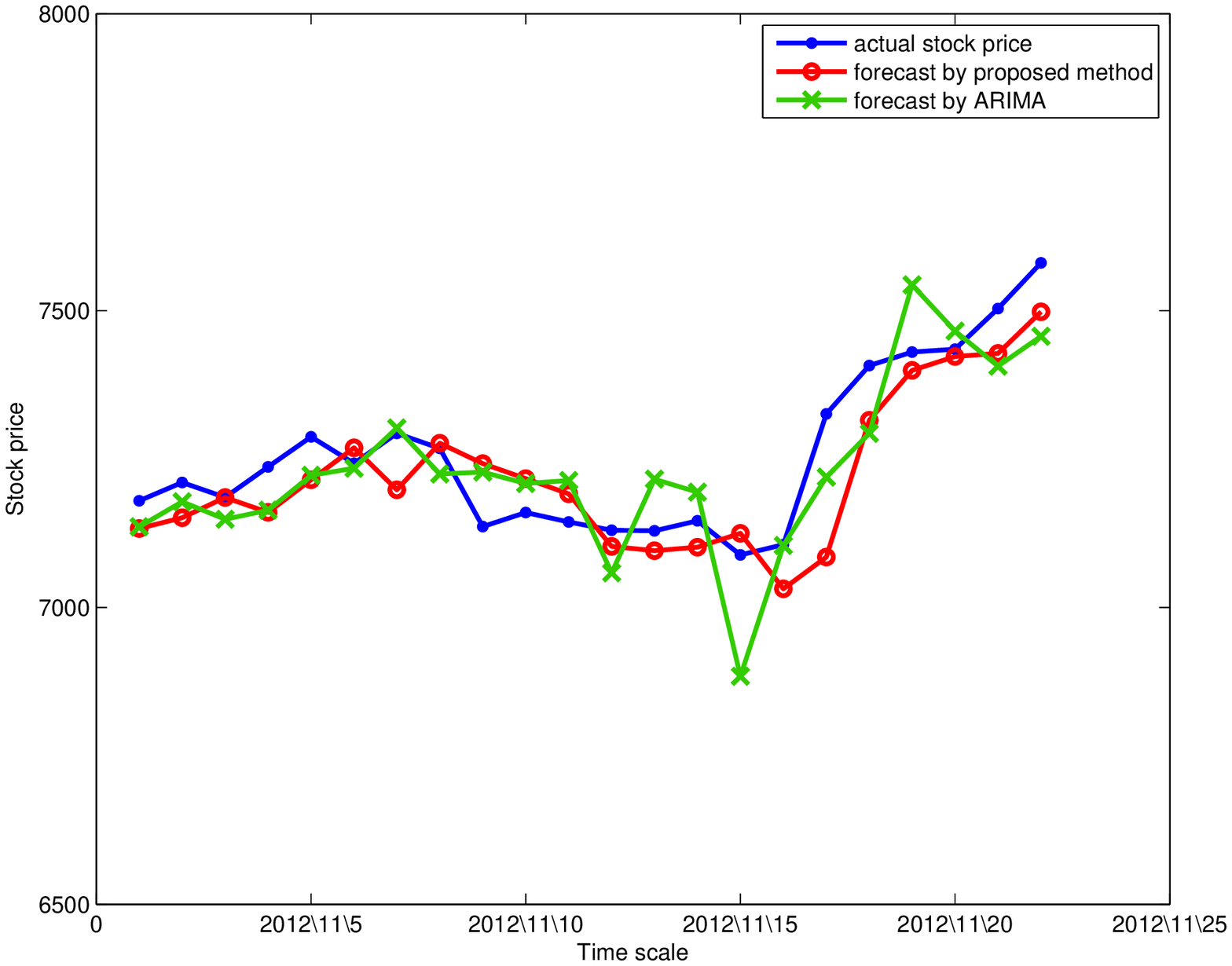}
\caption{\emph{The training data consists of the data of June, July, August, September and October ((107 records)). The figure shows the predicted stock price of November obtained by ARIMA and the proposed method, respectively.}}
\label{our5ARIMA5}
\end{figure}

The proposed method and ARIMA model use the training set to forecast the next datum value in the time series. With the prediction processing, the actual stock price will be added into the training data and the oldest datum will be deleted. The number of the training data keeps unchanged. The training data includes three groups. Group 1 originally consists of the data of October (21 records), group 2 originally consists of the data of August, September and October (64 records), and group 3 originally consists of June, July, August, September and October (107 records). These groups are respectively used to forecast the stock price values of November.  The predicted results are shown in the Fig.\ref{our1ARIMA1}, Fig.\ref{our3ARIMA3} and Fig.\ref{our5ARIMA5}. Table \ref{tableRMSE} shows the root mean square errors of ARIMA model and the proposed method.

%

\begin{table}[htbp]
\tiny
{\footnotesize%
\caption{The RMSE of ARIMA and the proposed method} \label{tableRMSE}
\begin{tabular*}{\columnwidth}{@{\extracolsep{\fill}}@{~~}ccc@{~~}}
\toprule%
Group & RMSE(ARIMA) & RMSE(the proposed method)
\\
\midrule
$1$ & $158.06$ & $116.00$
\\
$2$ & $103.78$ & $78.29$
\\
$3$ & $82.91$ & $78.15$
\\

\bottomrule
\end{tabular*}
}
\end{table}

\begin{figure}[!t]
\centering
\includegraphics[scale=0.4]{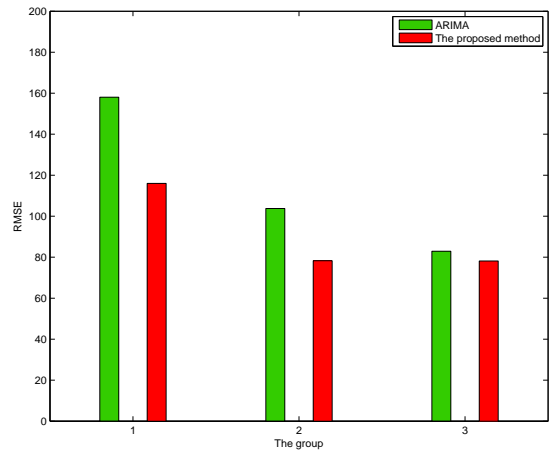}
\caption{\emph{This figure shows the RMSE of ARIMA and the proposed method. No matter what the number of training data is, the RMSE of the proposed method is less than ARIMA.}}
\label{RMSE}
\end{figure}

It is obvious that the forecasting performance of the proposed method is better than ARIMA model. Especially, when there is a small amount of data, the forecasting performance of the proposed method is much better than ARIMA model. With the number of data increasing, the predicted accuracy of the proposed method and ARIMA model is improved. The predicted results show that the proposed method is effective and when there is a small amount of data, the proposed method can obtain a better predicted performance than ARIMA model.

%
%

\section{Conclusion}

In this letter, a new forecasting method is proposed. This new method forecasts data from the aspect of the network. Time series is converted into network to find out which data will link with the datum which needs to be predicted. Then, the network which contains the relationship information between the predicted datum and other data is converted into time series again to forecast the value of datum which needs to be predicted. The case study illustrates that the proposed method is effectiveness. Compared with ARIMA, the proposed shows a good forecasting performance when there is a small amount of data.

\acknowledgments
The work described in this letter is partially supported Chongqing Natural
Science Foundation, Grant No. CSCT, 2010BA2003, National Natural
Science Foundation of China, Grant No. 61174022 and 71271061, National
High Technology Research and Development Program of China (863
Program) (No.2013AA013801), Science and Technology Planning Project
of Guangdong Province, China (2010B010600034, 2012B091100192),Doctor
Funding of Southwest University Grant No. SWU110021, Fundamental Research
Funds for the Central Universities No. XDJK2014D008.


\end{document}